\documentstyle[12pt]{article}
\newcommand{\be}{\begin{equation}}
\newcommand{\ee}{\end{equation}}
\newcommand{\ba}{\begin{eqnarray}}
\newcommand{\ea}{\end{eqnarray}}
\newcommand{\bi}[1]{\bibitem{#1}}
\newcommand{\fr}[2]{\frac{#1}{#2}}
\newcommand{\non}{\nonumber}
\newcommand{\ar}{\mbox{$\rightarrow$}}

\newcommand{\al}{\mbox{$\alpha$}}
\newcommand{\ka}{\mbox{$\kappa$}}
\newcommand{\val}{\mbox{$\vec{\alpha}$}}
\newcommand{\om}{\mbox{$\omega$}}
\newcommand{\Z}{\mbox{$Z\alpha$}}
\newcommand{\p}{\mbox{$\vec{p}$}}
\newcommand{\pp}{\mbox{$\vec{p}\,'$}}
\newcommand{\k}{\mbox{$\vec{k}$}}
\newcommand{\kp}{\mbox{$\vec{k'}$}}
\newcommand{\r}{\mbox{$\vec{r}$}}
\newcommand{\bt}{\mbox{$\beta$}}
\newcommand{\dE}{\mbox{$\Delta E$}}
\newcommand{\D}{\mbox{$\vec{D}$}}
\newcommand{\lb}{\left (}
\newcommand{\rb}{\right )}
\newcommand{\la}{\left\langle}
\newcommand{\ra}{\right\rangle}

\begin{document}
\pagestyle{empty}
\vspace{1.0cm}

\begin{center}
{\Large \bf Order $(Z\alpha)^4 \fr{m}{M} R_{\infty}$ Correction to
Hydrogen Levels}\\

\bigskip

{\bf A.S. Yelkhovsky}\\
Budker Institute of Nuclear Physics,\\
and \\
Physics Department, Novosibirsk University, \\
630090 Novosibirsk, Russia
\end{center}

\bigskip

\begin{abstract}
The first in $m/M$ and fourth in $Z\al$ (pure recoil) correction to a
hydrogen energy level is found. This correction comprises two contributions,
one coming from the atomic scale, the other from distances of the Compton
wavelength order. Two different perturbation schemes are used to calculate
the former. One of them exploits as unperturbed the solution to the
Dirac-Coulomb problem, the nucleus' slow motion being the source of the
perturbation. The alternative scheme treats both the electron and the
nucleus as slow, while relativistic effects are considered perturbatively.
The short-distance contribution is found in the Feynman and Coulomb gauges.
Recent results for $P$ levels are confirmed, in contrast with those for $S$
levels. Numerically, the shift equals 2.77 kHz for the ground state and 0.51
kHz for $2S$ state.
\end{abstract}

\newpage

\pagestyle{plain}
\pagenumbering{arabic}

\section{Introduction}

To determine the proton charge radius with a percent accuracy from the value
of the hydrogen Lamb shift, the latter should be known both experimentally
and theoretically with a precision of one kHz. Recently completed
calculation of the order $m\al^2(\Z)^5$ corrections \cite{E} leaves among
the effects of possible phenomenological interest the pure recoil correction
of the order $m^2(\Z)^6/M$ arising due to interference between the nucleus'
recoil and relativistic effects in the motion of the electron. The present
paper is devoted to the calculation of this correction for an arbitrary
state of the hydrogen atom.

Recently this correction for $P$ states was found \cite{p}. In those states,
as well as in all states with nonzero angular momenta, the correction proves
to be saturated by a contribution coming from the atomic scale. Hence one
can use there the standard quantum mechanical perturbation theory for the
effective operators describing relativistic effects. Matrix elements of
the effective operators arising in the perturbation theory converge at small
distances thus testifying {\it a posteriori} that the used 'nonrelativistic'
approach is correct at the given order of \al\ for states with nonzero $l$.

An attempt to apply the same approach to $S$ states, whose wavefunctions do
not vanish at the origin, leads to matrix elements diverging at small
distances. In fact, among the effective operators one finds those depending
on $r$ as $r^{-3}$ and even $r^{-4}$ \cite{p}. As for the latter, for $S$
states the operator $r^{-4}$ is equivalent (modulo a nonsingular operator)
to the sum of operators with the radial dependence $r^{-3}$ and
$\delta(\r)/r$. It was shown in Ref.\cite{FKMY} that logarithmically
divergent contributions are mutually cancelled. This cancellation means
that for the states with vanishing angular momentum, the correction we
discuss splits naturally into two contributions -- those of large and small
distances -- each gaining its value in its own scale. To calculate the
former, one can use again the nonrelativistic approach, whereas the short
distance contribution, residing in the Compton wavelength order scale, calls
for a true relativistic approach.

The closed expression for the first recoil correction to an energy of the
relativistic electron moving in the Coulomb field is outlined in Sec.2. This
expression is used in Sec.3 for the evaluation of the long-distance
contribution. It proves that the relativistic approach is more efficient even
at the atomic scale. The contribution of short distances is
found in Sec.4 employing the Feynman gauge. Sec.5 is devoted to checks of
the obtained results. The long-distance contribution is recalculated there
using the nonrelativistic approach, while the short-distance one is found in
the Coulomb gauge. Finally, in Sec.6, we give the numerical values for the
energy shifts and compare results of the present work with those obtained
earlier in \cite{p,PG}.

Throughout the paper the relativistic units $\hbar=c=1$ are used. Since we
do not discuss radiative corrections, $Z$ is also set equal to unity.

\section{Methods of Calculation}

One of the perturbation schemes we use at the present work starts from the
Schr\"odinger equation in the Coulomb field, when both particles are
considered as nonrelativistic in the zeroth approximation. To account for
relativistic effects, a kind of the operator product expansion is built by
calculation of scattering amplitudes for free relativistic particles. Thus
arising effective operators are then the subject for the ordinary
perturbation theory of the nonrelativistic quantum mechanics. This approach
is rather well suited for long-distance contributions, which are due to
effective operators saturated by nonrelativistic region and having non-local
kernels. Formerly it was used in calculations of i) logarithmic in \al\
corrections to the spectrum of the two-body system \cite{PhScr}, ii) the
order $m\al^6$ corrections to the positronium $P$ levels \cite{pos}, and
iii) the order $m^2\al^6/M$ corrections to the hydrogen $P$ levels \cite{p}.
Unfortunately, this approach becomes very tedious being applied to
short-distance contributions, when effective operators with local kernels
are represented by a number of diagrams.

An alternative approach deals with relativistic light particle (electron)
moving in the field generated by the slow heavy one (nucleus). In the zeroth
approximation, the heavy particle holds still being a source of the Coulomb
field.  Wavefunction of the system reduces to that of the light particle
satisfying the Dirac equation. To first order in the heavy particle's
inverse mass, the perturbation operator coincides with its nonrelativistic
Hamiltonian:
\be\label{Hh}
V = \fr{\lb \vec{P} - |e| \vec{A}(\vec{R}) \rb ^2}{2M}.
\ee
Here $\vec{P}$ is the operator of a nucleus' momentum. The vector potential
$\vec{A}$ acts at the nucleus' site.

Unfortunately, one cannot calculate an energy shift induced by the
perturbation (\ref{Hh}) straightforwardly, i.e. taking merely its
average. In fact, the operator (\ref{Hh}) depends on the nucleus' dynamical
variables while the argument of the unperturbed wavefunction is a position
or momentum of the electron. To overcome this difficulty we use the gauge
invariance of observables in the quantum electrodynamics \cite{Y}. Being
reexpressed in terms of electron's variables, the average value of
(\ref{Hh}) should be retained gauge invariant. The new form of the average
is now nearly evident:
\be\label{main}
\dE_{rec} = - \fr{1}{M} \int \fr{d\om}{2\pi i} \la \lb \p - \D \rb
            G_{E+\om} \lb \p - \D \rb \ra.
\ee
Here \p\ is the electron momentum operator, \D\ is the integral operator
describing the transverse quantum exchange. It has the kernel\footnote{In
what follows we will write often a kernel rather than an appropriate
operator for the sake of brevity.}
\[
 \fr{4\pi \al \val_k }{k^2 - \om^2}, \;\;\;\;\;
\val_k \equiv \val - \fr{\k (\val\k)}{k^2}.
\]
In (\ref{main}), $G$ is the Green's function for the Dirac equation in the
Coulomb field, the average is taken over a Dirac-Coulomb eigenstate with an
energy $E$.  Actually, the "seagull" part of (\ref{main}), that of the
second order in \D, emerges naturally as a counterpart of the $\vec{A}^2$
term from (\ref{Hh}) when we take the expectation value of this term over
fluctuations of the electromagnetic field. All the other terms provide the
invariance of (\ref{main}) with respect to the gauge transformation,
\be
\psi \ar \exp(i\phi(\r))\psi,\;\;\;\;\;\;\D \ar \D + i[\p,\phi].
\ee
One can easily convince oneself that just the same result can be obtained
from the formula (11) of \cite{Y} with the help of the Dirac equation.

The first attempt to obtain the relativistic expression for the recoil
correction to hydrogen energy levels was undertaken in Ref.\cite{Br}.
Complete expressions for various contributions in the Coulomb gauge were
originally derived in the framework of the quasipotential approach in
Ref.\cite{Sh}. The sum of those contributions can be convinced to reduce to
the right-hand side of (\ref{main}).

\section{Long-Distance Contribution}

Present section is devoted to the calculation of the contribution to the
energy shift which is saturated by the atomic scale and can thus be called
the long-distance one. To check the results, two approaches described above
have been applied in parallel. Here we describe in detail how the second,
relativistic, approach works. The procedure of comparison with the results
of the more cumbersome nonrelativistic approach will be postponed until the
Section 5.

\subsection{Pure Coulomb Contribution}

In the relativistic approach the pure Coulomb contribution,
\be
\dE_C = - \fr{1}{M} \int \fr{d\om}{2\pi i} \la \p G_{E+\om} \p \ra =
          \fr{1}{2M} \la \p \lb \Lambda_+ - \Lambda_- \rb \p \ra,
\ee
can naturally be represented as the sum of two terms \cite{Br},
\be\label{C}
\dE_C = \la \fr{p^2}{2M} \ra - \fr{1}{M} \la \p \Lambda_- \p \ra.
\ee
Here $\Lambda_+$ and $\Lambda_-$ are the projection operators to sets of
positive- and negative-energy Dirac-Coulomb eigenstates respectively.
With the aid of the Dirac equation the mean value of $p^2/2M$ can readily
be reexpressed in the following form \cite{Sh}:
\be\label{kin}
\la \fr{p^2}{2M} \ra = \fr{m^2-E^2}{2M} + \fr{m^2}{2M}
\la 2 \lb \fr{E}{m} -\bt \rb \fr{\al}{r} + \fr{\al^2}{r^2} \ra.
\ee
As for the second term in (\ref{C}), responsible for virtual transitions
into negative-energy states, the simple analysis shows that it doesn't
contribute to the order of interest at the atomic scale. Actually, the
trivial power counting on the right-hand side of the obvious inequality,
\be\label{in}
|\la \p \Lambda_- \p \ra| < \left|\fr{1}{4m^2}\la [\p,C]
     \Lambda_- [\p,C] \ra \right|,
\ee
where $C$ is the Coulomb potential, shows that at the atomic scale, the
product of commutators is already of the sixth order in \al, so that the
projector and the wavefunctions can sufficiently be replaced by their
nonrelativistic counterparts. Since there is no negative-energy states in
the nonrelativistic approximation, the atomic scale contribution to the
initial average also vanishes in the order we consider.

\subsection{Magnetic Contribution}

After performing the integration over \om, the expression for the single
transverse, or magnetic, contribution,
\be\label{M}
\dE_M = \fr{1}{M} \int \fr{d\om}{2\pi i} \la \p G \D + \D G \p \ra,
\ee
turns into
\be\label{Mld}
\dE_M = -\fr{\al}{M}\mbox{Re} \la \p \lb \sum_+
\fr{|m\rangle\langle m|}{k+E_m-E} - \sum_-\fr{|m\rangle\langle m|}{E-E_m+k}
\rb \fr{4\pi \val_k}{k} \ra,
\ee
where $\sum_+$ denotes the sum over discrete levels plus the integral over
positive-energy part of the continuous spectrum, while $\sum_-$ stands for
the integral over negative-energy continuum.

For the transverse photon momenta in the atomic region, $k \sim m\al$,
one can expand the first term in (\ref{Mld}) to the power series in the
ratio $(E-E_m)/k$. To zeroth order (in the approximation of the instant
exchange), we have:
\be\label{M0}
-\fr{\al}{M}\mbox{Re} \la \fr{4\pi \val_k}{k^2}\Lambda_+ \p \ra =
- \fr{m^2}{2M} \la 2 \lb \fr{E}{m} -\bt \rb \fr{\al}{r} +
\fr{\al^2}{r^2} \ra +\fr{\al}{M}\mbox{Re} \la
\fr{4\pi \val_k}{k^2}\Lambda_- \p \ra.
\ee
The sum of the first term and (\ref{kin}) has very simple form \cite{Sh},
\be\label{Inst}
\fr{m^2-E^2}{2M} = \fr{m^2\al^2}{2MN^2},
\ee
where the standard notations for the Dirac-Coulomb problem are used,
\[
N = \sqrt{(\gamma+n_r)^2 + \al^2}, \;\;\;\;\gamma = \sqrt{\ka^2-\al^2},
\]
$n_r$ is the radial quantum number, $\ka = - 1 - \vec{\sigma}\vec{l}$.
Notice that (\ref{Inst}) reduces to the lowest order result for the states
with $n_r=0$ only.

As far as we are seeking only for corrections of the even order in \al, the
next term of the expansion to be considered is
\be\label{M2}
\dE_{ret}=-\fr{\al}{M}\mbox{Re} \la \p \sum_+ (E_m-E)^2
|m\rangle\langle m| \fr{4\pi \val_k}{k^4} \ra = -\fr{\al}{M}\mbox{Re} \la
\left[ H, \left[ H, \p \right] \right] \Lambda_+ \fr{4\pi \val_k}{k^4} \ra,
\ee
where $H = \val\p + \bt m + C$ is the Dirac Hamiltonian in the Coulomb
field.  This term describes the effect of retardation. Implying the
corresponding operator by its kernel, we have
\[
[H,\p] = \al \fr{4\pi\k}{k^2}, \;\;\;\;\;
\left[H,[H,\p]\right] = \al \fr{4\pi\k(\val\k)}{k^2}.
\]
To the lowest nontrivial order, matrix elements of \val's over
positive-energy states can be replaced by the appropriate Pauli currents:
\be\label{pret}
\dE_{ret}\approx - \fr{\al^2}{M} \la \fr{4\pi\kp}{k'^2}
                  \fr{2\pp\kp-k'^2}{2m}\;\fr{4\pi}{k^4}
                  \fr{2\p_k+i\vec{\sigma}\times\k}{2m} \ra.
\ee
Here $\kp = \vec{q}-\k$, $\vec{q}=\pp-\p$, while \p\ and \pp\ are the
arguments of the wavefunction and its conjugated respectively. Being
converted to the spatial representation, the average above equals
\be\label{ret}
\dE_{ret} = \fr{\al^2}{4m^2M} \la -2\p\fr{1}{r^2}\p +
            \fr{7\vec{l}^2 + 2\vec{\sigma}\vec{l}}{2r^4} \ra.
\ee
Strictly speaking, in (\ref{pret}), the integral over \k\ has an infrared
divergent part. It is omitted from (\ref{ret}) since the photon momenta
$k\sim m\al^2$ contribute to the previous order correction. To make sure
that this is correct, one can regularize the divergency supplying the photon
with a mass $\lambda$ such that $m\al^2 \ll \lambda \ll m\al$. The term
proportional to $1/\lambda$ and omitted from (\ref{ret}) can be easily
checked to cancel the respective term in the difference between (\ref{Mld})
and the expression obtained from (\ref{Mld}) by the replacement $k \ar
\sqrt{k^2+\lambda^2}$.  On the other hand, just this difference determines
the low-energy contribution to the order $m\al^5/M$ correction.

The last of contributions due to the single transverse exchange is generated
by virtual transitions into negative-energy states and is covered by the
last terms in (\ref{Mld}) and (\ref{M0}). The inequality similar to
(\ref{in}) shows that the nonrelativistic expansion of the last term in
(\ref{Mld}) starts with the seventh power of \al. As for the negative-energy
contribution to (\ref{M0}), to the lowest nontrivial order it reduces to
\be\label{M-mom}
- \fr{\al^2}{2mM} \la \fr{4\pi \val_{k'}}{k'^2} \lambda_-(\p+\k)
             \fr{4\pi \k}{k^2} \ra \approx \fr{\al^2}{4m^2M} \la
             \fr{4\pi \val_{k'}\k}{k'^2} \fr{4\pi \val \k}{k^2} \ra,
\ee
yielding in the spatial representation
\be\label{M-}
\dE_{M-} = - \fr{\al^2}{4m^2M} \la \fr{1}{r^4} - \fr{4\pi\delta(\r)}{r} \ra.
\ee
Taken over $S$ states, this average is logarithmically divergent at small
distances (linear divergencies cancel each other). An ultraviolet divergency
will be discussed later for the {\em total} long-distance contribution to
the $S$ level shift. In fact, due to the gauge dependence of an individual
contribution (e. g., (\ref{M-})), its divergent part alone has no physical
meaning.

\subsection{Seagull Contribution}

Again, taking the integral over \om\ in the expression for the double
transverse, or seagull, contribution,
\be\label{S}
\dE_S = - \fr{1}{M} \int \fr{d\om}{2\pi i} \la \D G_{E+\om} \D \ra,
\ee
we obtain
\be\label{Sld}
\dE_S = \fr{\al^2}{2M} \la \fr{4\pi \val_{k'}}{k'} \sum_+
        \fr{|m\rangle\langle m|}{(E_m-E+k')(E_m-E+k)} \lb 1 +
        \fr{E_m-E}{k'+k}\rb \fr{4\pi \val_k}{k} + \cdots \ra,
\ee
where the ellipsis stands for the negative-energy part which differs from
the positive-energy one by the overall sign and signs before $k$ and $k'$.
One can easily check that the linear in $k/2m$ terms in the expansion of the
negative-energy part cancel each other. But just these terms at the atomic
scale could produce the energy correction of the necessary order. Hence it
remains to consider the positive-energy part explicitly written in
(\ref{Sld}). In the leading nonrelativistic approximation,
\be
\dE_{S+} = \fr{\al^2}{2M}\la\fr{4\pi\val_{k'}}{k'^2}\Lambda_+
          \fr{4\pi\val_k}{k^2}\ra,
\ee
we again replace matrix elements of \val's over positive-energy states by
the Pauli currents,
\be
\dE_{S+} = \fr{\al^2}{2M}\la\fr{4\pi}{k'^2}\fr{2\pp_{k'}+i\vec{\sigma}
           \times\kp}{2m}\fr{4\pi}{k^2}\fr{2\p_k+i\vec{\sigma}
           \times\k}{2m} \ra,
\ee
and perform the Fourier transformation to obtain
\be\label{S+}
\dE_{S+} = \fr{\al^2}{4m^2M} \la 2\p\fr{1}{r^2}\p + \fr{1}{r^4}
            - \fr{3\vec{l}^2 + 2\vec{\sigma}\vec{l}}{2r^4} \ra.
\ee

\subsection{Total Long-Distance Contribution}

Summing up (\ref{ret}), (\ref{M-}) and (\ref{S+}) we arrive at
\be\label{sum}
\fr{\al^2}{4m^2M} \la 2 \fr{\vec{l}^2}{r^4} + \fr{4\pi\delta(\r)}{r} \ra.
\ee

Expanding (\ref{Inst}) to the power series in $\al^2$ and evaluating the
average in (\ref{sum}) we obtain the long-distance contribution for a state
with nonzero $l$:
\ba\label{ld}
\dE_{l>0} &=& \fr{m^2\al^6}{Mn^3} \left\{ \fr{1}{8|\ka|^3} +
\fr{6\ka}{|\ka|(4\ka^2-1)(2\ka+3)} + \fr{3}{8n\ka^2} \right. \non \\
&&\left. - \fr{1}{n^2|\ka|} \lb 1 + 2\fr{\ka^2 (\ka+1)}{(4\ka^2-1)(2\ka+3)}
\rb + \fr{1}{2n^3} \right\}.
\ea
For $l=1$ it reproduces the result of the paper \cite{p}. As previously
mentioned, the effective operators of the order under consideration contain
singularities insufficient to compensate the vanishing of a wavefunction
with nonzero $l$ at $r\ar 0$. That is why (\ref{ld}) is the total sought-for
correction to $l>0$ levels.

For $S$ states the first term in (\ref{sum}) evidently vanishes due to the
angular momentum operator $\vec{l}$ annihilating their wavefunctions. It is
interesting to note that a na\"{\i}ve generalization of the result for
states with nonzero $l$ to $S$ ones leads to the error -- the vanishing of
the angular average is compensated by the linear divergency of the radial
one.

As for the second term, which formally contains the linear divergency, it is
just a remnant of the short-distance contribution to the previous (fifth in
\al) order correction. To make sure that nothing is lost in the sixth order,
let us regularize the ultraviolet divergency by subtraction of the potential
generated by the massive transverse exchange (the photon mass $\lambda \gg
m\al$), from the potential of the ordinary transverse exchange entering
(\ref{M-mom}):
\be\label{UVreg}
\fr{4\pi \val_{k'}}{k'^2} \;\; \ar \;\; \fr{4\pi \val_{k'}}{k'^2} -
\fr{4\pi \val_{k'}}{k'^2 + \lambda^2}.
\ee
By going to the spatial representation we obtain the regularized version of
the singular operator:
\[
\fr{4\pi\delta(\r)}{r} \;\; \ar \;\; \fr{2\lambda}{3} 4\pi\delta(\r).
\]
Being averaged, it gives the energy correction of the order
$m\lambda\al^5/M$. The latter should be cancelled by the linear in $\lambda$
term in the expansion of the short-distance contribution to the order
$m^2\al^5/M$ correction, calculated with the massive propagator of the
transverse quantum (actually the expansion parameter is $\lambda/m \ll 1$).
Along with linear in $\lambda/m$ correction, one could expect the correction
linear in $\al=m\al/m$. However the expansion parameter at large distances
is $(p/m)^2 \sim \al^2$ so that the operator we discuss does not contribute
to the order of interest. On the other hand, a linear in \al\ correction to
a local ($\propto \delta(\r)$) operator can arise as an ordinary radiative
one. In this case the correction is completely saturated by small distances.
The next section is devoted to the calculation of such corrections.

So, in $S$ states the long-distance contribution is exhausted by the
$m^2\al^6/M$ term from the expansion of (\ref{Inst}),
\be\label{ldS}
\dE_{l=0}^{ld} = \fr{m^2\al^6}{2Mn^3} \lb \fr{1}{4} + \fr{3}{4n} -
                 \fr{2}{n^2} + \fr{1}{n^3} \rb.
\ee
It vanishes in the ground state only.

\section{Short-Distance Contribution}

As we have mentioned in the Introduction, the long-distance contribution for
$S$ states is supplied by a short-distance one residing at scales of the
Compton wavelength order. Since two contributions are well separated, each
of them is gauge invariant, so that evaluating the short-distance one we can
use a different, more appropriate gauge. Rather naturally the mostly
convenient gauge is the Feynman one. The main formula (\ref{main}) can be
rewritten in this gauge by application of the Dirac equation or directly
from the eq.(11) of \cite{Y}:
\be\label{fey}
\dE_{tot} = - \fr{\al^2}{M} \int \fr{d\om}{2\pi i} \la
              \fr{4\pi}{k'^2 + \lambda^2 - \om^2}
              \lb \val + \fr{\kp}{\om} \rb G_{E+\om}
              \lb \val - \fr{\k}{\om} \rb
              \fr{4\pi}{k^2 + \lambda^2 - \om^2} \ra.
\ee
Momenta of the photons are assumed to flow both from the nucleus to the
electron. The photon mass $\lambda$ is introduced to establish control over
infrared divergences reminiscent of lower-order and long-distance
contributions.  Those divergencies arise in the process of the approximate
evaluation of the integrals in (\ref{fey}).

Taking the wavefunctions at the origin and replacing the Green's function by
the first term of its expansion in the Coulomb field, we have
\ba
\dE_G = \fr{\al^3\psi^2}{M}\int \fr{d\om}{2\pi i}\la\fr{4\pi}{p'^2-
        \sqrt{\;}^2}\lb\val-\fr{\p\,'}{\om}\rb
   \fr{m+\om+\bt m+\val\p\,'}{p'^2-\Omega^2}\fr{4\pi}{q^2}\right.&&\non\\
        \left.\fr{m+\om+\bt m+\val\p}{p^2-\Omega^2}\lb\val-\fr{\p}{\om}\rb
        \fr{4\pi}{p^2-\sqrt{\;}^2}\ra.&&
\ea
Here $\psi^2\equiv |\psi(0)|^2$, the angle brackets denote integrations over
$\p$ and $\p\,'$; $\vec{q} = \p\,'-\p$; and
\[
\sqrt{\;}\equiv \sqrt{\om^2-\lambda^2}, \;\;\;\;\;
\Omega\equiv\sqrt{2m\om+\om^2}.
\]
Contrary to the case of large distances, in the deep relativistic region the
opposite order of integration is suitable -- first over \p\ and \pp, and
then over \om. As for the former, it becomes rather trivial after conversion
to the spatial representation. Preparatory to such the conversion, it is
convenient to express all the scalar products containing different momenta
in terms of their squares. Then some of the denominators can be cancelled.
At this point we can drop those terms which do not contain $\Omega$ in their
denominators. In fact, the only scale leaving in such terms is $\lambda$ so
that they cannot produce a short-distance contribution. In the spatial
representation the initial two-loop integral with zero external momenta
turns into a simple one-dimensional integral over $r$. The contour
of the resulting \om-integration encloses the cut between the points $-2m$
and $-\lambda$ in the complex plane. After this last integration we obtain
\be\label{G}
\dE_G = \fr{\pi\al^3\psi^2}{Mm}\lb \fr{1}{\varepsilon} -
\fr{8}{3\pi\sqrt{\varepsilon}}\int_1^{\infty}\fr{dx}{\sqrt{x(x^2-1)}} +
4\ln 2 - \fr{5}{2} \rb,
\ee
where $\varepsilon=\lambda/2m$.

If the Green's function is taken to zeroth order in the interaction, we have
to use the Dirac equation in order to account for the momentum dependence in
the wavefunction:
\ba
\dE_{\psi} = 2\fr{\al^3\psi^2}{M}\int \fr{d\om}{2\pi i}\la\fr{4\pi}{p'^4}
        \lb 2m+\val\p\,'\rb\fr{4\pi}{q^2-\sqrt{\;}^2}
        \lb\val+\fr{\vec{q}}{\om}\rb\right.&&\non\\
        \left.\fr{m+\om+\bt m+\val\p}{p^2-\Omega^2}\lb\val-\fr{\p}{\om}\rb
        \fr{4\pi}{p^2-\sqrt{\;}^2}\ra.&&
\ea
Using the same procedure, we take $-r/2$ as the Fourier transform of
$4\pi/p^4$. The linearly divergent constant we thus leave aside is actually
proportional to $1/\al$ and contributes to the previous order correction.
The result of the integration is
\be\label{psi}
\dE_{\psi} = \fr{\pi\al^3\psi^2}{Mm}\lb -\fr{1}{2\varepsilon^2} +
\fr{1}{\varepsilon} -
\fr{8}{3\pi\sqrt{\varepsilon}}\int_1^{\infty}\fr{dx}{\sqrt{x(x^2-1)}}\rb.
\ee
Regulator-dependent terms in (\ref{G}) and (\ref{psi}) arise from the
integrals saturated by the region of momenta $p\sim\lambda$ and frequency
$\om\sim\lambda$ (or $\sqrt{m\lambda}$) and are thus the remnants of the
previous orders corrections or of the long-distance contribution. Truly
relativistic contribution comes from the region $p\sim\om\sim m$ and does
not depend on the infrared cutoff:
\be\label{sd}
\dE^{sd} = \fr{m^2\al^6}{Mn^3}\lb 4\ln 2 - \fr{5}{2} \rb \delta_{l0}.
\ee
It is pertinent to note here that this result is truly short-distance, i. e.
it does not contain hidden long-distance contributions, which na\"{\i}vely
could arise due to cancellation of the same nonzero powers of $\lambda$ from
numerator and denominator -- all positive powers of the photon mass were
dropped out in the process of calculation. On the other hand, an emergence
of such contributions would be self-contradictory. Actually, if an integral
is saturated by distances of $1/\lambda$ order, then at $p \sim \lambda$,
the integrand denominator has at least one power of momentum more than the
product of the numerator and the measure of integration. In other words, any
"long-distance" contribution (determined by the scale of $\lambda$) {\em
has} to contain a positive power of the photon mass in its denominator.

\section{Checking of the Results}

\subsection{Long-Distance Contribution}

To be certain that the long-distance contributions are found correctly, all
of them were rederived in the framework of the nonrelativistic approach
which exploits the Schr\"odinger equation as a starting point. For the
states with nonzero angular momenta we used the following procedure. All the
contributions prove to have the same analytic structure in \ka, namely
\[
\dE = \fr{m^2\al^6}{Mn^3} \Sigma,
\]
where
\ba
\Sigma &=& \fr{1}{|\ka|^3} \lb a_{\infty} + \fr{a_{1/2}}{\ka-1/2} +
             \fr{a_{-1/2}}{\ka+1/2} + \fr{a'_{-1/2}}{(\ka+1/2)^2}
            + \fr{a_{-1}}{\ka+1} + \fr{a_{-3/2}}{\ka+3/2} \rb \non \\
        && +\fr{1}{n\ka^2} \lb b_{\infty} + \fr{b_{-1/2}}{\ka+1/2} \rb \non
        \\ && + \fr{1}{n^2|\ka|} \lb c_{\infty} + \fr{c_{1/2}}{\ka-1/2} +
         \fr{c_{-1/2}}{\ka+1/2} + \fr{c_{-3/2}}{\ka+3/2} \rb+\fr{d}{n^3},
\ea
Constants $a,b,c,d$ evaluated in the nonrelativistic approach for
individual contributions as their asymptotic values at $\ka \ar \infty$ or
residues at corresponding poles were then compared with the respective
results obtained in the relativistic approach. In the process of comparison,
a number of 'nonrelativistic' contributions breaks down into groups
according to the meaning of respective 'relativistic' ones. For example, the
retardation part of the magnetic contribution (\ref{ret}) comprises three
terms in the nonrelativistic approach: $\dE_{MC}^{(1)}$, $\dE_{MCC}^{(1)}$
and $\dE_{ret}^{(1)}$ (notations are from ref.\cite{p}).

As we mentioned earlier, $S$ states should be treated separately in order to
avoid fictious contributions arising due to the compensation between
vanishing angular averages and linearly divergent radial ones. All the
ultraviolet divergencies in $S$ states are checked to cancel each other. To
this end we regularize the effective potentials which are too singular at $r
\ar 0$ and ensure that the total long-distance contribution for $S$ states
is independent of the regularization parameter.

\subsection{Short-Distance Contribution}

In order to compare the results of the present work with those of \cite{PG},
the short-distance contribution was calculated using the Coulomb gauge also.
A mass of the magnetic quantum was used as the infrared regulator. The
scheme of calculation is completely analogous to those used previously in
the case of the Feynman gauge. The short-distance contributions are:
\be\label{Csd}
C_G = C_{\psi} = \fr{\pi\al^3\psi^2}{Mm} \fr{1}{2};
\ee
\medskip
\ba
M_G &=& - \fr{\pi\al^3\psi^2}{Mm} \lb \ln\fr{\lambda}{2m}+\fr{3}{2} \rb, \\
M_{\psi} &=& - \fr{\pi\al^3\psi^2}{Mm} \ln\fr{\lambda}{2m};
\ea
\medskip
\ba
S_G &=& \fr{\pi\al^3\psi^2}{Mm} \lb 4\ln 2 - 2 \rb, \\
S_{\psi} &=& \fr{\pi\al^3\psi^2}{Mm} 2 \ln\fr{\lambda}{2m}.\label{Ssd}
\ea
Here $C$, $M$, $S$ denote Coulomb, magnetic and seagull contributions
respectively. It is easy to check that the sum of these contributions
coincides with (\ref{sd}).

\section{Conclusion}

Numerically, the correction to the energy equals 2.77 kHz for the ground
state and 0.51 kHz for the $2S$ state. Being somewhat less than the
na\"{\i}ve estimate ($m^2\al^6/M \approx 10.2$ kHz) it nevertheless is quite
comparable with the accuracy of the near future measurements. It is also
interesting to note that the corrections to $2S$ and $2P$ levels (with the
radiative-recoil correction to $2P$ level \cite{p} taken into account) are
rather close to each other, so that the correction to their difference, 0.04
kHz, can be considered as negligibly small at the present level of the
experimental accuracy \cite{exp}.

Let us now set up a correspondence between the results of the present work
and those of the other papers. The result for $l>0$ levels appears to be
firmly established \cite{p,ASY}. For $S$ levels our result is contradictory
to the recent results of the analytic \cite{PG} and numerical \cite{ASY}
calculations.

To elucidate the origin of the disagreement, we consider the correction to
the ground state energy. It is easy to verify that our short-distance
results (\ref{Csd})--(\ref{Ssd}) are in one-to-one correspondence with
respective "high-energy" contributions from \cite{PG}. A similar statement
is true for long-distance contributions (low- and intermediate-energy ones
in notations of \cite{PG}), with one exception. The coefficient --2 from
Eq.(68) of Ref.\cite{PG} for the intermediate-energy contribution to the
retarded exchange by the magnetic quantum, differs from our result, --1 (in
the same units $m^2\al^6/M$), which arises after trivial averaging in
(\ref{ret}) over the ground state. Unfortunately, we have not managed to
reproduce the coefficient --2 starting from Eq.(67) of Ref.(\ref{ret}).
Furthermore, several arguments can be brought forward, that the result
(68),\cite{PG} for the retardation contribution looks at least suspicious.
In particular, the logarithmic divergency in the order $m^2\al^6/M$ is known
to appear due to the relativistic corrections to the {\em instant}
transverse exchange.  The result of the present work concerning the origin
of this logarithmic divergency and the value of the corresponding
coefficient is contained in (\ref{M-}) and is in complete agreement with
those of Refs.\cite{PhScr} and \cite{DGE}. As for the effect of retardation,
it gives rise to the finite contribution only (in accord with (\ref{ret})).
But it follows from the result of Ref.\cite{PG} that just the retardation is
the source of not only logarithmic, but even the linear divergency at small
distances, while the long-distance relativistic correction to the instant
transverse exchange does not contribute at all.

\bigskip

{\bf Acknowledgments}\nopagebreak

The author is grateful to A. Vainshtein for stimulating discussions, and to
M. Eides, S. Karshenboim, and V. Shabaev for their interest to the work.
The work was supported by a fellowship of INTAS Grant 93-2492 and
is carried out within the research program of International Center for
Fundamental Physics in Moscow. Partial support from the program
``Universities of Russia'', Grant 94-6.7-2053, is also gratefully
acknowledged.

\end{document}